\documentclass[aps,prl,twocolumn,superscriptaddress]{revtex4}
\usepackage{graphics,epsfig}

\begin{document}


\title{Exploiting environmental resonances to enhance qubit quality factors}

\author{Silvia Kleff}
\email[]{kleff@theorie.physik.uni-muenchen.de}
\affiliation{Sektion Physik and CeNS, Ludwig-Maximilians Universit{\"a}t, Theresienstr. 37, 80333 M{\"u}nchen, Germany }
\author{Stefan Kehrein}
\affiliation{Theoretische Physik III, Universit{\"a}t Augsburg, 86135 Augsburg, Germany}
\author{Jan von Delft}
\affiliation{Sektion Physik and CeNS, Ludwig-Maximilians Universit{\"a}t, Theresienstr. 37, 80333 M{\"u}nchen, Germany }


\date{\today}

\begin{abstract}
We discuss dephasing times for a two-level system (including bias) coupled to a damped harmonic oscillator. 
This system is realized in measurements on solid-state Josephson qubits. It can be mapped to a 
spin-boson model with a spectral function with an approximately Lorentzian resonance. 
We diagonalize the model by means 
of infinitesimal unitary transformations (flow equations), 
and calculate correlation functions, dephasing rates, and qubit quality factors. 
We find that these depend strongly on the environmental resonance frequency $\Omega$; in particular, 
quality factors can be  enhanced significantly
by  tuning $\Omega$ to lie \emph{below} the qubit frequency $\Delta$. 
  
\end{abstract}


\maketitle



\textit{Introduction.--}
A key feature in qubit design is to gain
good control of dephasing induced by
the environment. A much-studied model
that has yielded considerable insight
into the dephasing of qubits (more generally,
2-state systems) is the spin-boson model~\cite{weiss}.
Most studies of this model assume a spectral
function $J(\omega)$ that has a power law form.
However, several qubit systems of current interest are coupled
to an environment which features rather strong
resonances, which would correspond to
a spectral function $J(\omega)$ with well-defined
peaks at characteristic frequencies.
A prominent example is the case of flux-qubits~\cite{wal:00} which are
read out using a SQUID, which has a characteristic
resonance frequency $\Omega$ (of order $3$ GHz),
which  in order of magnitude is comparable to 
the characteristic qubit energy scales ($10$ GHz)~\cite{wal:03,chiorescu:03}.

The presence of environmental resonances raises
several interesting questions with both fundamental
and practical implications: How is the qubit dynamics influenced by the
presence of environmental resonances? Can
the latter be used to indirectly tune qubit properties
such as the tunneling rate or q-factor?
Is it more advantageous  to have the resonance frequency
higher or lower than the characteristic qubit energies?

In this Letter, we explore these questions
in the framework of a model that has been
used with great success to describe and optimize
recent generations of flux qubits \cite{wal:03}: it involves
a spin degree of freedom (qubit) coupled to
a harmonic oscillator with frequency $\Omega$ (modeling the environmental
resonance), which in turn is coupled to a bath
of harmonic oscillators (to provide damping)~\cite{other}. 
It can be mapped \cite{garg:85}
onto a regular spin-boson model with a spectral
function $J(\omega)$ featuring an almost Lorentzian resonance peak near $\Omega$.
We are interested
not only in the regime where the qubit tunneling rate $\Delta$ is
much smaller than $\Omega$ (which would correspond
to the standard spin-boson model, with $\Omega$ playing
the role of the bath cut-off frequency) but also in 
the hitherto unexplored regime $\Delta > \Omega$. Here standard weak-coupling, poor man
scaling approach that predicts a downward renormalization
of $\Delta$ is insufficient; instead, we need
a method sufficiently powerful to deal
with all ratios of $\Delta/ \Omega$. To this end,
we use the flow-equation renormalization group (FER) approach of Wegner~\cite{wegner:94}
and of Glazek and Wilson~\cite{glazek:93}. Interestingly, we find that $\Delta$
is renormalized \emph{upwards} if the initial $\Delta$ is greater than
$\Omega$, and that correspondingly, the dephasing times and q-factors
are strongly \emph{increased}. These results have the
very important implication that \emph{by appropriately tuning the
environmental resonance frequency $\Omega$,
significant additional  control of the qubit dynamics can indeed
by obtained.}

\textit{Spin-boson-model.--} We consider the Hamiltonian
\begin{eqnarray}
       \tilde{{\mathcal H}} &=& -\frac{\Delta}{2}\sigma_{\mathrm{x}}+\frac{\varepsilon}{2}\sigma_{\mathrm{z}}+ (B^{\dagger}+B)\left[g\sigma_{\mathrm{z}}+\sum_{\mathrm{k}}\kappa_{\mathrm{k}}^{\phantom{\dagger}}(\tilde{b}_{\mathrm{k}}^{\dagger}+\tilde{b}_{\mathrm{k}}^{\phantom{\dagger}})\right]\nonumber\\
                          &+&\Omega B^{\dagger}B   + \sum_{\mathrm{k}}\tilde{\omega}_{\mathrm{k}}^{\phantom{\dagger}}\tilde{b}_{\mathrm{k}}^{\dagger}\tilde{b}_{\mathrm{k}}^{\phantom{\dagger}}                     
                           + (B^{\dagger}+B)^2\sum_{\mathrm{k}}\frac{\kappa_{\mathrm{k}}^{2}}
{\tilde{\omega}_{\mathrm{k}}},
\label{eq:system}
   \end{eqnarray}
which describes a 2-state-system with asymmetry energy $\varepsilon$ and 
tunneling matrix element $\Delta$, coupled linearly with 
strength $g$ to a harmonic oscillator with frequency $\Omega$, which is itself 
linearly coupled with strengths $\kappa_{\mathrm{k}}$ to a bath of harmonic oscillators.
The coupling to the environment is completely defined by
 the spectral function 
$\tilde{J}(\omega)\equiv\sum_{\mathrm{k}}\kappa_{\mathrm{k}}^{2}\delta(\omega-\tilde{\omega}_{\mathrm{k}})
=\Gamma\omega\Theta(\omega_c-\omega)$, which is as usual taken to be of ohmic form to model the dissipative environment. 
This system can be mapped to a spin-boson model of the form~\cite{garg:85}
   \begin{equation} 
       {\mathcal H}=-\frac{\Delta}{2}\sigma_{\mathrm{x}} +\frac{\varepsilon}{2}\sigma_{\mathrm{z}}
                    +\frac{1}{2}\sigma_{\mathrm{z}}\sum_{\mathrm{k}}\lambda_{\mathrm{k}}^{\phantom{\dagger}} (b_{\mathrm{k}}^{\dagger}+b_{\mathrm{k}}^{\phantom{\dagger}})
                    +\sum_{\mathrm{k}}\omega_{\mathrm{k}}^{\phantom{\dagger}}b_{\mathrm{k}}^{\dagger}b_{\mathrm{k}}^{\phantom{\dagger}},\nonumber
   \label{eq:spin_boson}
   \end{equation}
where spin dynamics depends only on the  structured spectral function
$J(\omega)\equiv\sum_{\mathrm{k}}\lambda_{\mathrm{k}}^{2}\delta(\omega-\omega_{\mathrm{k}})$ 
given by~\cite{wal:03}
   \begin{equation} 
       J(\omega)=\frac{2\alpha\omega\Omega^4\Theta(\omega_c-\omega)}{(\Omega^2-\omega^2)^2+(2\pi \Gamma\omega\Omega)^2},
                 \textrm{ with }\alpha=\frac{8\Gamma g^2}{\Omega^2}.
   \label{eq:density}
   \end{equation}
\textit{Flow equation renormalization.--}
Flow equation renormalization (FER) is based on infinitesimal 
unitary transformations of the Hamiltonian~\cite{wegner:94}. 
We follow the approach of Ref.~\cite{kehrein:97}, and mention only the main steps here:

(a) In order to decouple the two-level system from its environment we apply a sequence of 
\textit{unitary transformations} $U(l)$ to Eq.(\ref{eq:spin_boson}): $\mathcal{H}(l)=U(l)\mathcal{H}U^\dagger (l)$.
Here $\mathcal{H}(l=0)=\mathcal{H}$ is the initial Hamiltonian, $\mathcal{H}(l=\infty)$
is the final, diagonal Hamiltonian and $l$ denotes the flow parameter, which characterizes the square of the 
inverse energy scale being decoupled. In differential formulation this transformation reads 
\begin{equation}
   \label{eq:differentiell}
  \frac{d{\mathcal H}(l)}{dl}=[\eta(l),{\mathcal H}(l)]\quad\textrm{with}\quad
\eta (l)=\frac{dU(l)}{dl}U^{-1}(l). 
\end{equation}

(b) The canonical choice for the   \textit{generator} $\eta$ 
suggested by Wegner is  $\eta_c =[\mathcal{H}_0,\mathcal{H}]$ 
with $H_0=-\frac{\Delta}{2}\sigma_{\mathrm{x}}+\frac{\varepsilon}{2}\sigma_{\mathrm{z}}+
\sum_{\mathrm{k}}\omega_{\mathrm{k}}^{\phantom{\dagger}}
b_{\mathrm{k}}^{\dagger}b_{\mathrm{k}}^{\phantom{\dagger}}$~\cite{wegner:94}.
    However, since $\eta_c$ generates new coupling terms (originally not present in the Hamiltonian) 
    it is advisable to modify our generator. For $\varepsilon\neq 0$ we choose~\cite{zerobias}
\begin{eqnarray}
\label{eq:generator}
\eta & = & i\sigma_{\mathrm{y}}\sum_{\mathrm{k}}\eta^{\mathrm{y}}_{\mathrm{k}}(b_{\mathrm{k}}^{\phantom{\dagger}}+b_{\mathrm{k}}^\dagger)
+\sigma_{\mathrm{z}}\sum_{\mathrm{k}}\eta_{\mathrm{k}}^{\mathrm{z}}(b_{\mathrm{k}}^{\phantom{\dagger}}- b_{\mathrm{k}}^\dagger)\\\nonumber
& + & \sigma_x\sum_{\mathrm{k}}\eta_{\mathrm{k}}^{\mathrm{x}}(b_{\mathrm{k}}^{\phantom{\dagger}}-b_{\mathrm{k}}^\dagger)
+\sum_{\mathrm{kq}}\eta_{\mathrm{kq}}^{\phantom{\dagger}}(b_{\mathrm{k}}^{\phantom{\dagger}}+b_{\mathrm{k}}^\dagger)(b_q^{\phantom{\dagger}}-b_q^\dagger),
\end{eqnarray}
for which Eq.(\ref{eq:differentiell}) closes for terms linear in bosonic operators.
We neglect small higher-order terms in $[\eta,\mathcal{H}]$ that contain a coupling of the system to two bosonic modes.
The parameters $\eta_{\mathrm{k}}$ and $\eta_{\mathrm{kq}}$ in Eq.(\ref{eq:generator}) are given by $\eta^{\mathrm{x}}_{\mathrm{k}}=-(\lambda_{\mathrm{k}}/2)(\varepsilon\Delta /\omega_{\mathrm{k}})f(\omega_{\mathrm{k}},l)$, $\eta^{\mathrm{y}}_{\mathrm{k}}=-(\lambda_{\mathrm{k}}/2)\Delta f(\omega_{\mathrm{k}},l)$, $\eta^{\mathrm{z}}_{\mathrm{k}}=-(\lambda_{\mathrm{k}}/2)[(\omega^2_{\mathrm{k}}-\varepsilon^2) /\omega_{\mathrm{k}}]f(\omega_{\mathrm{k}},l)$, and 
$\eta_{\mathrm{kq}}=\Delta^2/(2\Delta_{\varepsilon})
\tanh{(\beta\Delta_{\varepsilon}/2)}
\lambda_{\mathrm{k}}\lambda_{\mathrm{q}}\omega_{\mathrm{q}}/(\omega_{\mathrm{k}}^2-\omega_{\mathrm{q}}^2)
[f(\omega_{\mathrm{k}},l)+f(\omega_{\mathrm{q}},l)]$ with 
$\Delta_{\varepsilon}=\sqrt{\Delta^2+\varepsilon^2}$.
 We choose $f(\omega_{\mathrm{k}},l)=\left[\omega^2(\omega_{\mathrm{k}}-\Delta_\varepsilon)\right]/\left[\Delta_\varepsilon^2(\omega_{\mathrm{k}}+\Delta_\varepsilon)\right]$. 
By comparing numerical results for the $\varepsilon=0$ (see~\cite{zerobias}) and the $\varepsilon\neq 0$ Ansatz
 we see that for 
$\varepsilon\neq 0$, due to our particular choice of $f(\omega_{\mathrm{k}},l)$,  we are restricted to couplings $\alpha\lesssim0.02$ which is a reasonable bound for experimental realizations. 
For an alternative Ansatz (for $\varepsilon\neq 0$) see also Ref.~\cite{stauber:02}.

(c) Eqs.(\ref{eq:differentiell}) and (\ref{eq:generator}) give us a set of differential equations (\textrm{flow equations}) for the parameters
in the Hamiltonian, namely $\varepsilon (l)$, $\Delta (l)$, and $\lambda_{\mathrm{k}}(l)$ [respectively $J(\omega ,l)$]:
\begin{eqnarray}
\label{eq:flow1}
-\partial_l\Delta/\Delta&=&\!\int\!\! d\omega\coth{\left[\frac{\beta\omega}{2}\right]} J(\omega ,l)f(\omega ,l),\quad\!\partial_l\varepsilon=0, \\\nonumber
\partial_l J(\omega,l)&=&-2f(\omega ,l)(\omega^2-{\Delta_{\varepsilon}}^2)J(\omega,l)+\tanh{\frac{\beta\Delta_{\varepsilon}}{2}}\times\\
\frac{2\Delta^2}{\Delta_{\varepsilon}}\hspace{-0.2cm}&J&\hspace{-0.2cm}(\omega ,l)
 \int d\omega^\prime
 \frac{\omega^\prime J(\omega^\prime ,l)}{\omega^2-{{\omega}^\prime}^2}
[f(\omega ,l)+f(\omega^\prime ,l)].
\end{eqnarray}
Note that according to (\ref{eq:flow1}) the bias $\varepsilon$ is not renormalized. 
The renormalization of the bath frequencies $\omega_{\mathrm{k}}$ vanishes in the thermodynamic limit of infinitely many modes.

(d) \textit{Observables} such as $\sigma_{\mathrm{z}}$ have 
to be subject to the same sequence of infinitesimal transformations 
as the Hamiltonian: $d{\sigma_{\mathrm{z}}}(l)/dl=[\eta(l),\sigma_{\mathrm{z}}(l)]$. 
For the flow of $\sigma_{\mathrm{z}}$ we make the Ansatz~\cite{zerobias}
\begin{eqnarray}
\label{eq:sigma}
\sigma_{\mathrm{z}}(l)&=&h(l)\sigma_{\mathrm{z}}+ s(l)\sigma_{\mathrm{x}}+r(l)+i\sigma_{\mathrm{y}}\sum_{\mathrm{k}}\mu_{\mathrm{k}}^{\mathrm{y}}(l)(b_{\mathrm{k}}^{\phantom{\dagger}}-b_{\mathrm{k}}^\dagger)\nonumber\\
&+&\sum_{\mathrm{k}}\left[\sigma_{\mathrm{x}}\chi_{\mathrm{k}}^{\mathrm{x}}(l)+\sigma_{\mathrm{z}}\chi_{\mathrm{k}}^{\mathrm{z}}(l)\right](b_{\mathrm{k}}^{\phantom{\dagger}}+b_{\mathrm{k}}^\dagger)
\end{eqnarray}
We neglect (small) terms in $[\eta,\sigma_{\mathrm{z}}]$ which contain a coupling to two bosonic modes.
The calculation of the flow equations for the six parameters $h$, $s$, $r$, $\chi_{\mathrm{k}}^{\mathrm{x}}$, 
$\chi_{\mathrm{k}}^{\mathrm{z}}$, and $\mu_{\mathrm{k}}^{\mathrm{y}}$ in Eq.(\ref{eq:sigma}) is straightforward~\cite{kleff:03}.

(e) To calculate \textit{correlation functions} of the form 
$C(t)=\frac{1}{2}\langle \sigma_{\mathrm{z}}(t)\sigma_{\mathrm{z}}(0)
+ \sigma_{\mathrm{z}}(0)\sigma_{\mathrm{z}}(t)\rangle$
we use  
the decoupled Hamiltonian
${\mathcal H(\infty )}=-[\Delta(\infty)/2]\sigma_{\mathrm{x}}+[\varepsilon/2]\sigma_{\mathrm{z}}
                    +\sum_{\mathrm{k}}\omega_{\mathrm{k}}^{\phantom{\dagger}}b_{\mathrm{k}}^{\dagger}b_{\mathrm{k}}^{\phantom{\dagger}}$,
  and Eq.(\ref{eq:sigma}) for $l=\infty$. For $T=0$  the Fourier transform $C(\omega)$ of $C(t)$ then takes the form 
(here all parameters are taken at $l=\infty$):~\cite{zerobias}
\begin{figure}[t]
\begin{center}
  {\includegraphics[clip,width=0.75\linewidth]{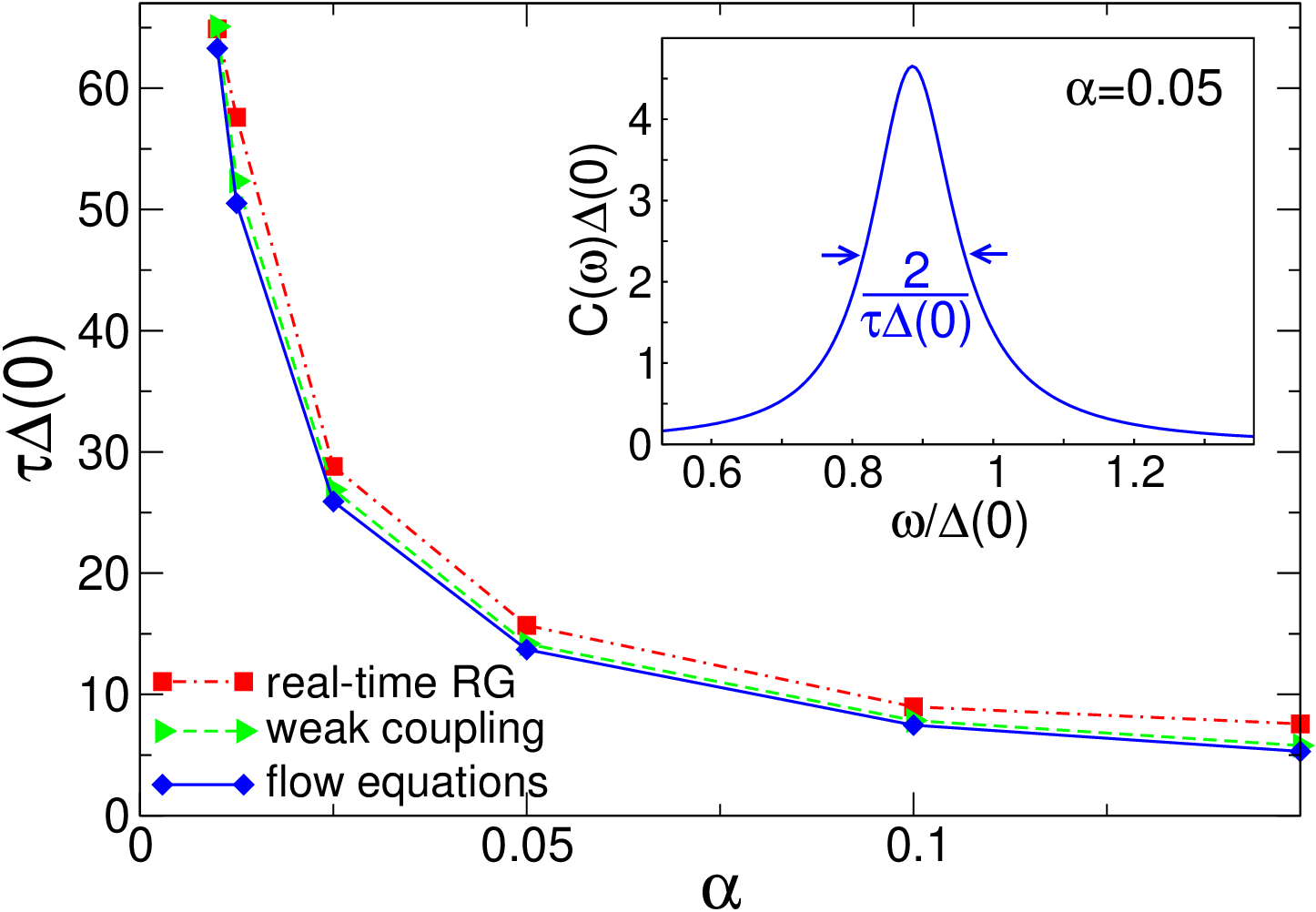}} 
\caption{Dephasing times for an ohmic bath with spectral function $J(\omega)=2\alpha\omega\Theta (\omega_c -\omega)$ as a function of 
$\alpha$. The FER result [$\varepsilon =0$ and $\omega_c=10\Delta(0)$] is 
compared with results from RTRG calculations~\cite{keil:01} and 
WCC~\cite{grifoni:99}. The inset shows a typical FER spin-spin correlation 
function.\vspace{-0.5cm}}
\label{fig:ohmic}
\end{center}
\end{figure}
\begin{eqnarray}
C(\omega )&=&\hspace{-0.1cm}\left[\frac{\varepsilon s}{\Delta_{\varepsilon}}+
\frac{\Delta h}{\Delta_{\varepsilon}}\right]^2
\hspace{-0.1cm}\delta (\omega-\Delta_{\varepsilon})+\left[\frac{\varepsilon h}{\Delta_{\varepsilon}}-
\frac{\Delta s}{\Delta_{\varepsilon}}-r\right]^2
\hspace{-0.1cm}\delta (\omega)\nonumber\\
&+&\hspace{-0.1cm}\sum_{\mathrm{k}}
\left[\frac{\varepsilon}{\Delta_{\varepsilon}}\chi_{\mathrm{k}}^{\mathrm{x}}-\mu_{\mathrm{k}}^{\mathrm{y}}
+\frac{\Delta}{\Delta_{\varepsilon}}\chi_{\mathrm{k}}^{\mathrm{z}}\right]^2
\delta(\omega-[\omega_{\mathrm{k}}+\Delta_{\varepsilon}])\nonumber\\
&+&\hspace{-0.1cm}\sum_{\mathrm{k}}\left[\frac{\varepsilon}{\Delta_{\varepsilon}}\chi_{\mathrm{k}}^{\mathrm{z}}-
\frac{\Delta}{\Delta_{\varepsilon}}\chi_{\mathrm{k}}^{\mathrm{x}}\right]^2
\delta (\omega-\omega_{\mathrm{k}}).
\label{eq:correlation}
\end{eqnarray} 
Numerically, one
finds that $h(\infty)=s(\infty)=0$ and i) $r(\infty)=0$ for
$\epsilon=0$ or ii) $r(\infty) \neq 0$ for $\epsilon \neq 0$.
Therefore, of the terms in the first line of~(\ref{eq:correlation}) only $\delta(\omega)$
remains and describes the nonzero expectation value of
$\sigma_{\mathrm{z}}$ for systems with asymmetry.

 In order to obtain quantitative results for the correlation function $C(\omega)$, we  numerically 
integrate the flow
equations  up to some value $l_0$, which is taken sufficiently large that the final 
results do not  depend on  it.
\begin{figure}[t]
\begin{center}
      {\includegraphics[clip,width=0.855\linewidth]{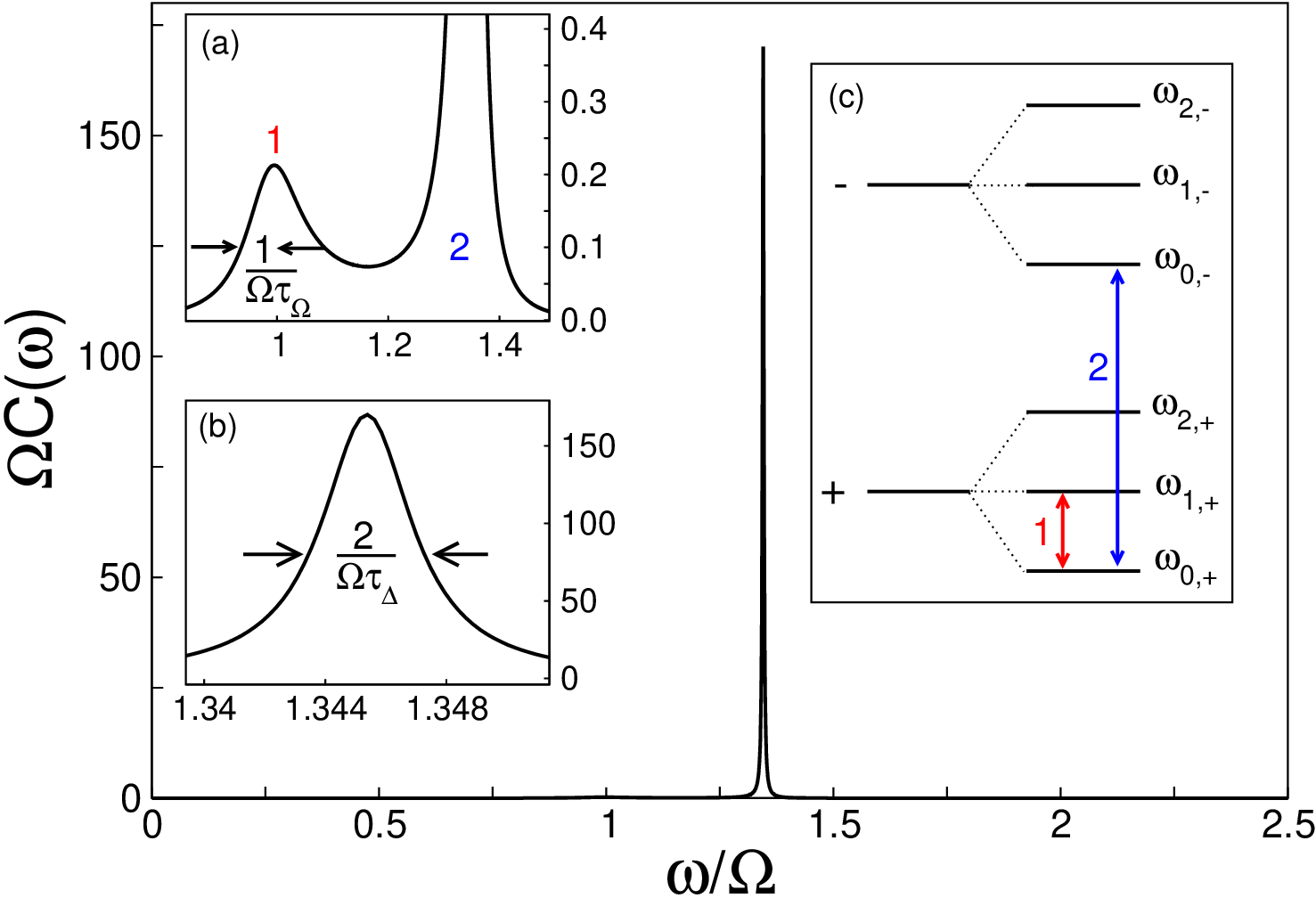}} 
\caption{Spin-spin correlation function as a function of frequency for
 experimentally relevant parameters discussed in Ref.~\cite{wal:03}:
$\alpha =0.0006$, $\Delta(0)=4$ GHz, $\varepsilon =0$ (this is the so-called ``idle state''), $\Omega =3$ GHz, 
$\Gamma =0.02$, and $\omega_c=8$ GHz. 
The sum rule is fulfilled with an error of less than $1$\%.
(a) Blow up of the peak region reveals a double peak; (b) blow up of the larger peak, 
(c) term scheme of a two level system coupled to an harmonic oscillator, drawn for $\Delta(0)\gg\Omega$;
($\alpha=0.0006$ corresponds to $g/\Omega\approx 0.06$.)\vspace{-0.9cm}}
\label{fig:correlation_weak}
\end{center}
\end{figure}
From the numerical results for  $C(\omega)$, which reflects the dynamics of the two-level system, we extract ``dephasing
times'', defined as the widths at half maximum of the resonances occurring in  $C(\omega)$ (as depicted in the inset of 
Fig.~\ref{fig:ohmic}). For zero bias ($\varepsilon =0$) a sum rule of the form 
$\int_0^\infty d\omega C(\omega )=1$ should hold~\cite{kehrein:97}.

\textit{Ohmic bath:} 
We start by  comparing our results for the dephasing time 
for a spin-boson model with an ohmic bath, $J(\omega)=2\alpha\omega\Theta(\omega_c-\omega)$, with results from 
real-time renormalization group (RTRG)~\cite{keil:01} 
and weak coupling calculations (WCC)~\cite{grifoni:99}.
Figure~\ref{fig:ohmic} shows the dephasing time $\tau$ 
as a function of the coupling strength $\alpha$.
 For weak 
coupling the dephasing time (at $T=0$) is given by~\cite{grifoni:99}
\begin{equation} 
\tau_{\mathrm{w}}=4/J[\Delta_\varepsilon (\infty)].
\label{eq:wcc}
\end{equation}
We find very good agreement with RTRG and WCC.

\textit{Structured bath/weak coupling:} We now turn to the structured spectral density given by Eq.(\ref{eq:density}). 
The main features of the corresponding system [Eq.(\ref{eq:system})] can already be understood by
analyzing only the coupled two-level-harmonic oscillator
system (without damping, i.e. $\Gamma=0$). For $\varepsilon=0$ this system 
exhibits two characteristic frequencies, close to  $\Omega$ and $\Delta$, associated with the transitions 
$1$ and $2$ in Fig.~\ref{fig:correlation_weak}(c).
These should also show up in the correlation function $C(\omega)$; and indeed   
Fig.~\ref{fig:correlation_weak}(a)   displays a \textit{double-peak} structure with the peak separation 
somewhat larger than $(\Delta-\Omega)$, due to level repulsion. The coupling to the bath 
will in general lead to a broadening of the resonances and an enhancement of the repulsion of the two 
energies. Due to the very small coupling ($\alpha=0.0006$)
peak positions of $C(\omega)$ in Fig.~\ref{fig:correlation_weak} can with very good accuracy be derived from a  
second order perturbation calculation for the coupled two-level-harmonic oscillator
system, yielding the following transition  frequencies [depicted in
inset (c) of Fig.~\ref{fig:correlation_weak}]: 
$\omega_{1,+}-\omega_{0,+}=\Omega-g^22\Delta(0)/[\Delta^2(0)-\Omega^2]\approx 0.987\Omega$ and 
$\omega_{0,-}-\omega_{0,+}=\Delta(0)+g^22\Delta(0)/[\Delta^2(0)-\Omega^2]\approx 1.346\Omega$. 
With the two peaks we associate two different dephasing times, $\tau_\Omega$ and $\tau_\Delta$, as shown in 
 inset (a) and (b) of Fig.~\ref{fig:correlation_weak}. 
\begin{figure}[t]
\begin{center}
      {\includegraphics[clip,width=0.9\linewidth]{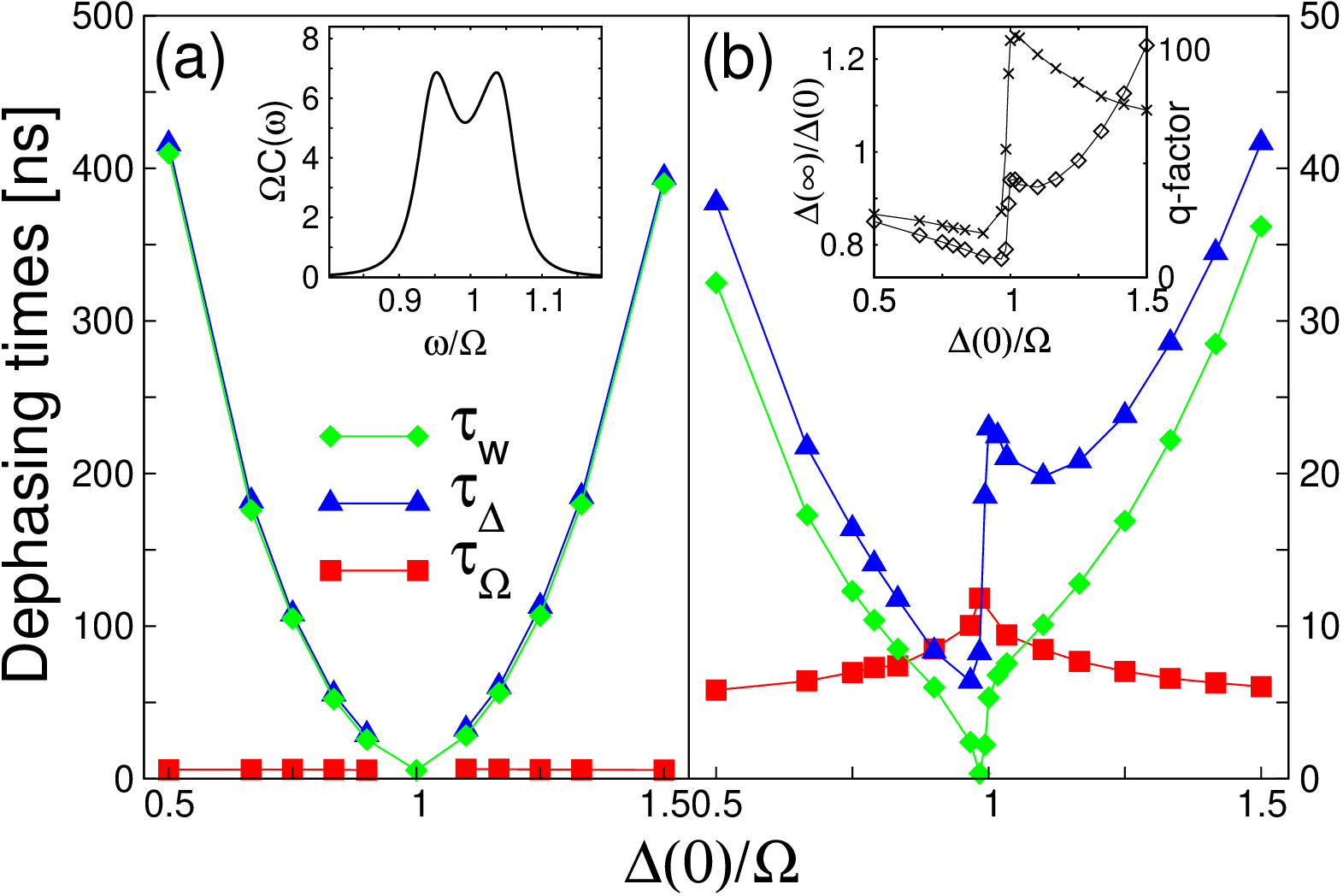}} 
\caption{FER results for dephasing times ($\tau_\Delta$ and $\tau_\Omega$)
for the structured bath [Eq.(\ref{eq:density})] 
compared to results from WCC ($\tau_{\mathrm{w}}$) given by Eq.(\ref{eq:wcc}), with
$\Delta(\infty)$ occurring therein calculated using FER. 
Parameters are
 $\varepsilon =0$, $\Omega =3$ GHz, 
$\Gamma =0.02$.  (a) weak coupling: $\alpha =0.0006$ and $\omega_c=8$ GHz. 
Sum rules are fulfilled with errors of less than $1$\%. Inset: Spin-spin correlation function for 
$\Delta(0)=0.987\Omega$. (b) stronger coupling: $\alpha =0.01$ and $\omega_c=9$ GHz.
Sum rules are fulfilled with errors of less than $3$\%. Inset: $\times$ - Renormalized tunneling 
matrix element; $\diamond$ - quality factor $q=\tau_\Delta\Delta(\infty)/2$.\vspace{-0.9cm}}
\label{fig:times_nobias}
\end{center}
\end{figure}
In Fig.~\ref{fig:times_nobias}(a) both these dephasing times
are shown as  functions of $\Delta(0)/\Omega$ for $\alpha=0.0006$. 
Moreover, $\tau_\Delta$ is compared to the WCC result $\tau_{\mathrm{w}}$ of  Eq.(\ref{eq:wcc}).
This comparison is expected to work well for $J(\Omega)/\Omega\ll1$, and indeed it does 
[here $J(\Omega)/\Omega\approx 0.08$]. 
 Due to the small coupling between two-level-system and harmonic oscilltor,
$g\approx 0.06\Omega$, the dependence of $\tau_\Omega$  
on $\Delta(0)$ is very weak. 

For $\Delta(0)$ close to $\Omega$ the two resonances in $C(\omega)$ merge to a 
symmetric double-peak structure as shown in the inset of Fig.~\ref{fig:times_nobias}(a). Here a
characterization by two different time scales becomes difficult. Therefore the corresponding data
points in Fig.~\ref{fig:times_nobias}(a) have not been included.

\begin{figure}[t]
\begin{center}
      {\includegraphics[clip,width=0.795\linewidth]{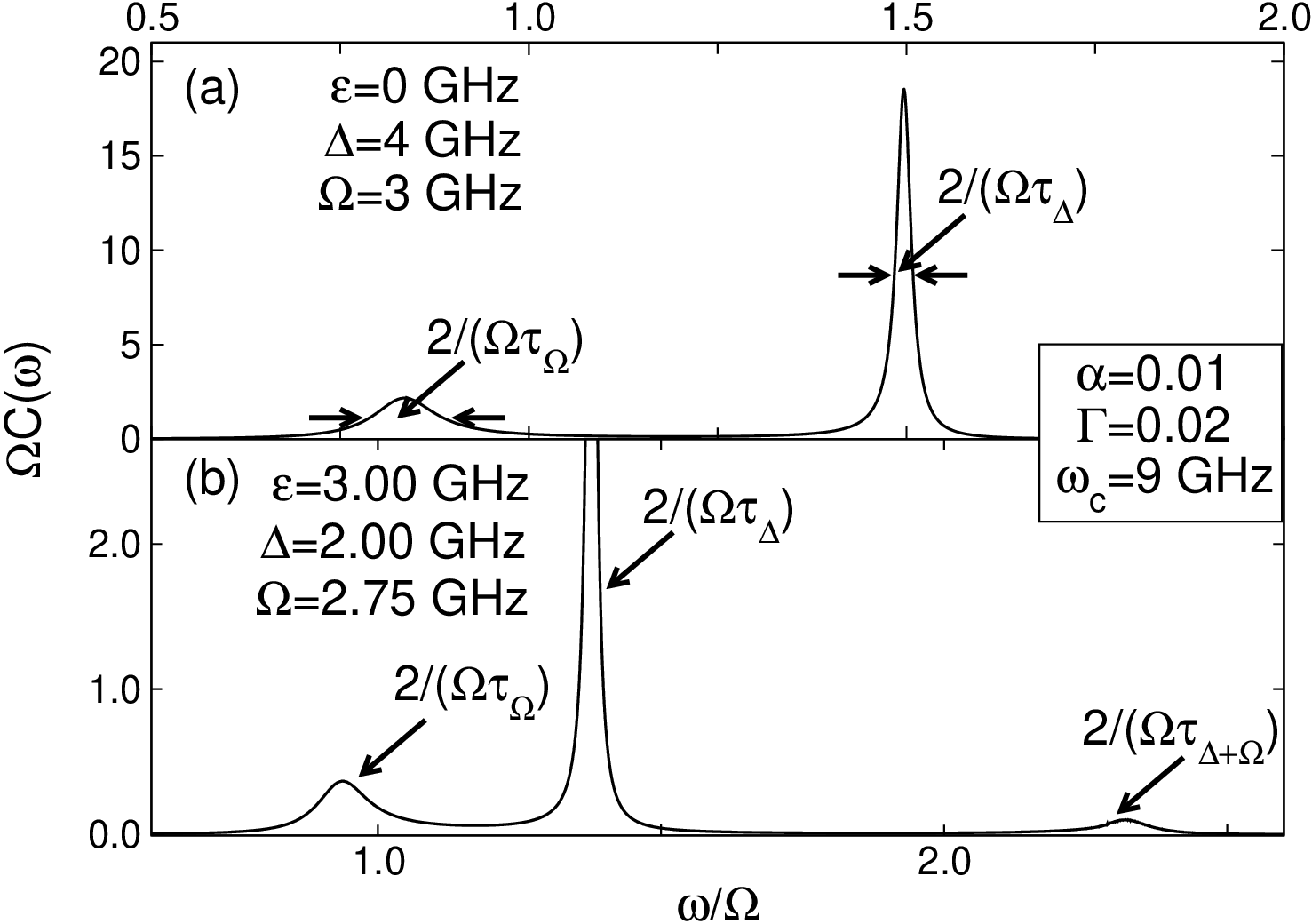}} 
\caption{Spin-spin correlation function for the structured bath [Eq.(\ref{eq:density})]
as a function of frequency~\cite{note1}. The maximum height of the middle peak in (b) is $\approx 7.2$.\vspace{-0.9cm}}
\label{fig:correlation_strong}
\end{center}
\end{figure}

\textit{Stronger coupling to bath:}
Figure~\ref{fig:times_nobias}(b) shows $\tau_\Delta$, $\tau_\Omega$, and $\tau_{\mathrm{w}}$ for a
larger coupling strength of $\alpha=0.01$. Figure~\ref{fig:correlation_strong}(a) shows one of the calculated
correlation functions. Note that the stronger coupling $\alpha$
leads to a larger separation, or ``level repulsion'', between  the $\Delta$- and  $\Omega$-peaks than in  
Fig.~\ref{fig:correlation_weak}. The inset of Fig.~\ref{fig:times_nobias}(b) shows the 
renormalized tunneling matrix element $\Delta(\infty)$ as a function the initial matrix element
$\Delta(0)$. Very importantly, for $\Delta(0)\gtrsim\Omega$, $\Delta$ increases during the flow, whereas for 
$\Delta(0)\lesssim\Omega$, it decreases~\cite{kleff:03_jpsj}.
This behavior can be understood from the fact that $f(\omega,l)$ in Eq.(\ref{eq:flow1}) changes sign at $\omega=\Delta$:
If the weight of $J(\omega)$ under  the integral in~(\ref{eq:flow1}) is larger for $\omega >\Delta(0)$, which is the case 
if $\Delta <\Omega$, then positive values of 
$f(\omega,l)$ dominate and $\Delta(l)$ decreases [$\Delta(\infty)<\Delta(0)$]; conversely 
if the weight is larger for $\omega<\Delta(0)$, $\Delta(l)$ increases [$\Delta(\infty)>\Delta(0)$].
Note also, that the upward renormalization towards larger $\Delta(\infty)$ in the inset of 
Fig.~\ref{fig:times_nobias}(b) is stronger than the downward one 
towards smaller values, i.e.,
the renormalization is \textit{not symmetric} with respect to $\Delta(0)=\Omega$. 
The reason for this \textit{asymmetry} lies in the fact that $f(\omega,l)$ has a larger weight for 
$\omega<\Delta$ than for $\omega>\Delta$.   
Also $\tau_\Delta$   and even $\tau_{\mathrm{w}}=1/J[\Delta(\infty)]$ in Fig.~\ref{fig:times_nobias}(b) show an asymmetric behavior 
with a steep increase at $\Delta(0)\approx\Omega$: dephasing times for $\Delta(0)>\Omega$ are larger than for  
$\Delta(0)<\Omega$. That this happens, although $J$ is more or less symmetric around
its maximum,  is a direct consequence of the stronger renormalization of $\Delta$ for the
latter case. Also the quality factor [$q$-factor, see inset in Fig.~\ref{fig:times_nobias}(b)] defined as 
$q=\tau_\Delta\Delta(\infty)/2$ shows this asymmetric behavior
 (being larger for $\Delta(0)>\Omega$ than for $\Delta(0)<\Omega$) with a steep increase at $\Delta(0)=\Omega$ 
from $8$ to $43$, i.e.,  
by a remarkably large factor of $\approx 5$. We consider the asymmetry of the renormalized tunneling matrix element, 
the dephasing time and the quality factor as the central results of this Letter: 
By tuning $\Omega$  such that
$\Delta(0)>\Omega$, dephasing times can be significantly  enhanced  (as compared to $\Delta(0)<\Omega$).  
Note also that $\tau_\Omega$ in Fig.~\ref{fig:times_nobias}(b) shows a 
much stronger dependence on 
$\Delta(0)$ than in Fig.~\ref{fig:times_nobias}(a). This is due to the stronger coupling  ($g\approx 0.3\Omega$)
of the 
two-level system to the harmonic oscillator in (b).

\begin{figure}[t]
\begin{center}
      {\includegraphics[clip,width=0.85\linewidth]{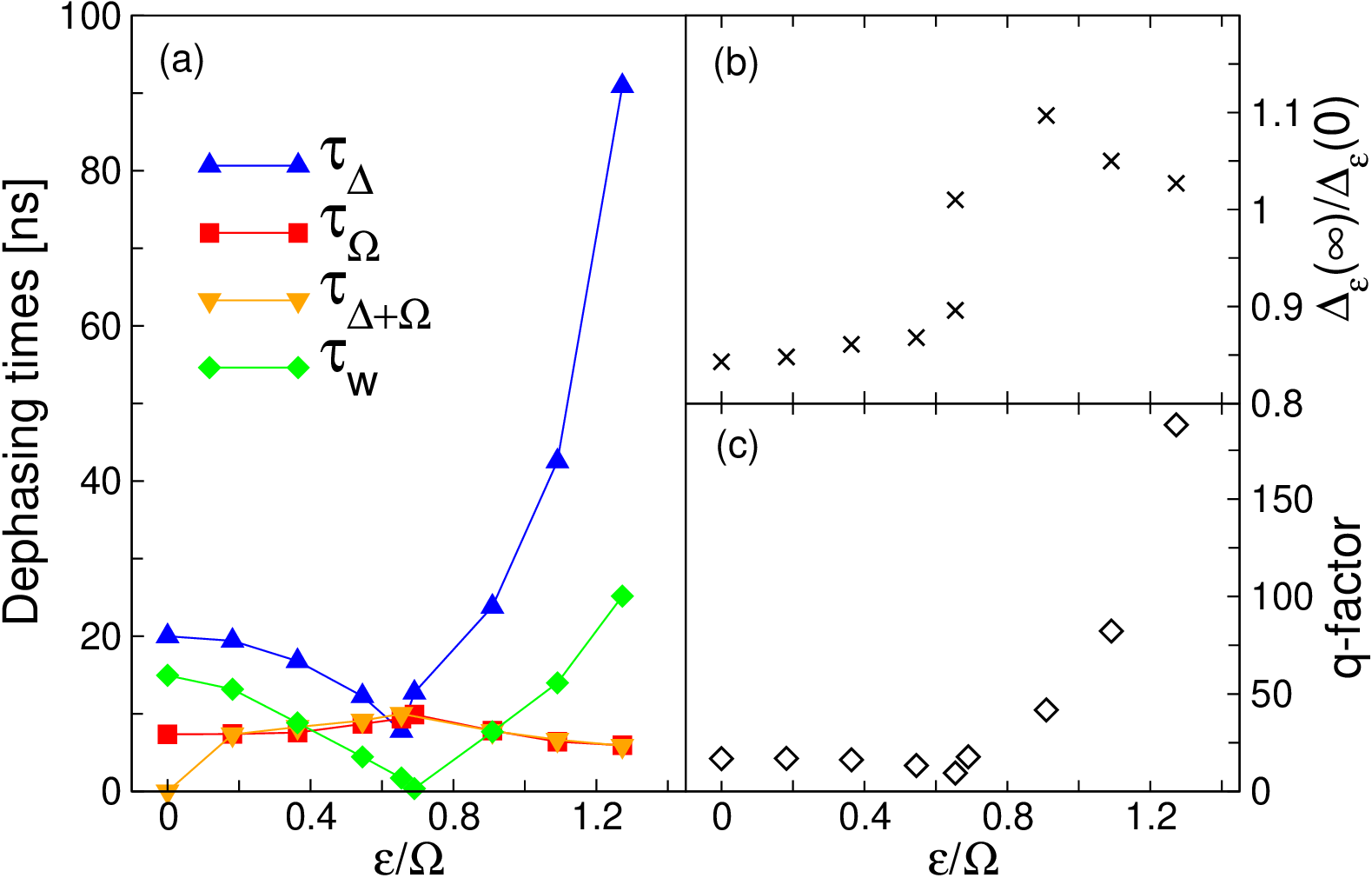}} 
\caption{(a) FER results for dephasing times ($\tau_\Delta$, $\tau_\Omega$, and $\tau_{\Delta +\Omega}$) 
for the structured bath [Eq.(\ref{eq:density})], 
compared to results from WCC ($\tau_{\mathrm{w}}$)
given by Eq.(\ref{eq:wcc}). (b) Renormalized tunneling matrix element. 
(c) q-factor $q=\tau_\Delta\Delta_\varepsilon(\infty)/2$. 
Parameters:
 $\alpha =0.01$, $\Delta(0)=2$ GHz, $\Omega =2.75$ GHz,  
$\Gamma =0.02$, and $\omega_c=9$ GHz.\vspace{-0.9cm}}
\label{fig:times_bias}
\end{center}
\end{figure}

\textit{Nonzero bias:} We now turn to the case of nonzero bias, $\varepsilon\neq 0$. A second order 
perturbation calculation, analogous to the zero bias case,
shows that a third resonance in $C(\omega)$ is expected to show up at an energy scale $\Delta_\varepsilon+\Omega$. 
Indeed it does, as exemplified in Figure~\ref{fig:correlation_strong}(b), which  shows a typical result for $C(\omega)$ for nonzero bias~\cite{note1}.
With  every resonance we associate a dephasing time 
(analogous to the zero bias case). 
Figure~\ref{fig:times_bias}(a) shows all three dephasing times ($\tau_\Delta$, $\tau_\Omega$, and $\tau_{\Delta+\Omega}$)
as a function of $\varepsilon$.
$\tau_\Delta$ is compared  to the weak coupling result $\tau_{\mathrm{w}}$. 
As expected, $\tau_\Delta$ shows a minimum at 
$\varepsilon_{\mathrm{min}}\approx\sqrt{\Omega^2-\Delta^2(0)}$, which corresponds
to the maximum of $J(\omega,l=0)$. Beyond $\varepsilon_{\mathrm{min}}$, $\tau_\Delta$ increases 
whereas $\tau_\Omega$ and $\tau_{\Delta+\Omega}$ decrease. This is
the expected behavior: In the limit $\varepsilon\to\infty$ ($\Delta\to 0$), $C(t)$ 
respectively $C(\omega)$ should become independent of all bath characteristics, i.e.,
 $C(t)\to 1$ and  $C(\omega)\to\delta(\omega)$. In this limit the dephasing times
should show the following behavior:  
$\tau_\Delta\to\infty$, $\tau_\Omega\to 0$ and $\tau_{\Delta+\Omega}\to 0$.
In Fig.~\ref{fig:times_bias}(b) and (c) the renormalized tunneling matrix element and the quality factor are shown as
function of the bias. Note that since $\varepsilon$ is not renormalized [see Eq.(\ref{eq:flow1})], 
$\Delta_\varepsilon (\infty)$ as a function of $\varepsilon/\Omega$ does not show a strong asymmetry, in contrast to the case  $\varepsilon =0$. As a direct consequence, 
dephasing times and quality factors do not change 
much at $\varepsilon_{\mathrm{min}}$.  Finally, $\tau$ and $q$
as a function of  $\Delta (0)$ for fixed $\varepsilon$ can be shown~\cite{kleff:03} 
to show a qualitatively similar behavior to 
Fig.~\ref{fig:times_nobias}.  

\textit{Summary --} We used flow equation renormalization to study 
a 2-level-system coupled to a damped harmonic oscillator for arbitrary ratios of $\Delta/\Omega$. 
We find that by tuning the system into the regime $\Delta>\Omega$, which is studied here for the first time,
dephasing times and q-factors can be significantly enhanced.

\begin{acknowledgments}
The authors are very grateful to F. Wilhelm and M.~Thorwart for interesting  discussions. 
S. Kehrein acknowledges support by
the SFB 484 of the Deutsche Forschungsgemeinschaft.
\end{acknowledgments}


\begin{thebibliography}{11}
\bibitem{zerobias}
Ans{\"a}tze and results for zero bias ($\varepsilon=0$) (see Ref~\cite{kehrein:97})
do not follow directly from the nonzero bias case. We therefore recall them here:
The generator in Ref.~\cite{kehrein:97} differs from Eq.(\ref{eq:generator}) through the absence of the term
proportional to $\sigma_{\mathrm{x}}$ and $f(\omega,l)$ was chosen as 
$f(\omega_{\mathrm{k}},l)=(\omega_{\mathrm{k}}-\Delta)/(\omega_{\mathrm{k}}+\Delta)$.
The truncation scheme for Eq.(\ref{eq:differentiell}) for this choice of $\eta$ 
works with good accuracy  for couplings $\alpha\lesssim 0.2$~\cite{kehrein:97}.
For $\varepsilon =0$ the Ansatz for $\sigma_{\mathrm{z}}$ reads: 
$\sigma_{\mathrm{z}}(l)=h(l)\sigma_{\mathrm{z}}+\sigma_{\mathrm{x}}\sum_{\mathrm{k}}\chi_{\mathrm{k}}^{\mathrm{x}}(l)
(b_{\mathrm{k}}^{\phantom{\dagger}}+b_{\mathrm{k}}^{\dagger})$,
which yields a spin-spin correlation function of the form 
$C(\omega)=\sum_{\mathrm{k}}\left[\chi_{\mathrm{k}}^{\mathrm{x}}\right]^2
\delta (\omega-\omega_{\mathrm{k}})$.
\bibitem{note1} Only contributions from the second and third line of Eq.(\ref{eq:correlation}) are shown in 
Fig.~\ref{fig:correlation_strong}(b). The $\delta$-function of the first line 
correspond to a constant part of $C(t)$.
\end{thebibliography}


\end{document}